%% file: BEP_DMA_arxiv.tex
\documentclass[twocolumn,conference]{IEEEtran}
\usepackage[T1]{fontenc}
\usepackage[latin9]{inputenc}
\usepackage{amsmath}
\usepackage{amsthm}
\usepackage{amssymb}
\usepackage{stmaryrd}
\usepackage{graphicx}
\usepackage[unicode=true,
 bookmarks=true,bookmarksnumbered=true,bookmarksopen=true,bookmarksopenlevel=1,
 breaklinks=false,pdfborder={0 0 0},pdfborderstyle={},backref=false,colorlinks=false]
 {hyperref}
\hypersetup{pdftitle={Your Title},
 pdfauthor={Your Name},
 pdfpagelayout=OneColumn, pdfnewwindow=true, pdfstartview=XYZ, plainpages=false}

\makeatletter
\theoremstyle{plain}
\newtheorem{thm}{\protect\theoremname}
\theoremstyle{plain}
\newtheorem{lem}[thm]{\protect\lemmaname}

\usepackage[caption=false,font=footnotesize]{subfig}

\DeclareMathOperator{\minimize}{minimize}
\DeclareMathOperator{\st}{subject~to}

\DeclareMathOperator{\tr}{Tr}

\DeclareMathOperator{\vect}{vec}

\usepackage{cite}
\usepackage{acronym}
\AtBeginDocument{\input{acro.tex}}

\makeatother

\providecommand{\lemmaname}{Lemma}
\providecommand{\theoremname}{Theorem}

\begin{document}
\title{On the Bit Error Probability of DMA-Based Systems\vspace{-3mm}}
\author{\IEEEauthorblockN{Nemanja~Stefan~Perovi\'c,~\emph{Member},~\emph{IEEE}}}
\maketitle
\begin{abstract}
Dynamic metasurface antennas (DMAs) are an alternative application
of metasurfaces as active reconfigurable antennas with advanced analog
signal processing and beamforming capabilities, which have been proposed
to replace conventional antenna arrays for next generation transceivers.
Motivated by this, we investigate the bit error probability (BEP)
optimization in a DMA-based system, propose an iterative optimization
algorithm, which adjusts the transmit precoder and the weights of
the DMA elements, prove its convergence and derive complexity. \acresetall{}
\end{abstract}


\section{Introduction}

\bstctlcite{BSTcontrol}While some of the advanced wireless technologies,
such as \ac{mMIMO} and \ac{mmWave} communications, aimed to enable
ubiquitous wireless connectivity and enhance network capacity, their
applications are inherently associated with increased hardware complexity
and excessively high energy consumption. The need to include aspects
of (energy) sustainability in the development of the future wireless
networks, as well as recent results in the field of configurable antennas
and metasurfaces have led to a fundamental change in the wireless
networks design paradigm \cite{di2020smart}. It implies exploring
new physical dimensions and applying radically new technologies for
the physical layer.

In previous years, metamaterials have emerged as a powerful technology
with a broad range of applications. Metamaterials comprise a variety
of artificial materials whose physical properties, and particularly
their permittivity and permeability, can be engineered to exhibit
various desired characteristics. A metasurface is an ultra-thin planar
structure engineered by compactly arranging a large number of electrically
controllable metamaterial elements, whose effective parameters can
be tailored to realize a desired transformation on the transmitted,
received, or impinging EM waves. Metasurfaces whose functions can
be modified after being manufactured and deployed are defined as dynamic
metasurfaces \cite{di2022communication}. Two main types of dynamically
tuned metasurfaces have been considered to date in the context of
wireless communications: passive reflective surfaces and active antenna
arrays. In previous years, passive reflective surfaces, known as \acp{RIS},
have been more widely researched. Broadly speaking, RISs are metasurfaces
whose reflection characteristics can be programmed by the utilization
of nearly passive integrated electronic circuits, in order to dynamically
manipulate incoming electromagnetic waves and control their propagation,
suppress interference, and enhance signals, in a wide variety of functionalities.
On the other hand, \acp{DMA} have been proposed as an efficient realization
of massive antenna arrays. They are based on the application of metasurfaces
as active reconfigurable antennas, and flexible architectures with
a massive amount of elements in a limited surface area and a reduced
number of RF chains, which facilitate signal processing in the analog
domain, decrease hardware complexity and power consumption \cite{shlezinger2021dynamic}.

The concept of \acp{DMA} in wireless communications was for the first
time introduced in \cite{shlezinger2019dynamic}, where \acp{DMA}
were used instead of \acp{mMIMO} in uplink communications. In addition,
the authors presented two algorithms for designing practical DMAs,
which provide the achievable sum-rates comparable to those in conventional
mMIMO systems. The maximization of the energy efficiency for the same
communication scenario was studied in \cite{you2022energy}. For downlink
multi-user communications, an algorithm that maximizes the achievable
sum-rates by adjusting the transmit precoder and the DMA weights was
proposed in \cite{wang2019dynamic}. The achievable rate optimization
for a DMA-based point-to-point MIMO system was studied in \cite{kimaryo2024uplink}.

As there are no papers in the literature addressing the \ac{BEP}
of DMA-based systems, we consider the BEP optimization of such systems
in this paper. More precisely, we formulate a joint optimization problem
of the transmit precoding and the wights of the DMA radiating elements
to minimize the BEP. To solve this problem, we propose an iterative
\ac{PGM}, for which the exact gradient and projection expressions
are presented in closed form. Furthermore, we prove the convergence
of the proposed algorithm. In addition, we provide the computational
complexity of the proposed algorithm \ac{wrt} the number of complex
multiplications. Through simulations, we show that the BEP can significantly
change with a number of microstrips in DMA-based systems with a larger
size of symbol alphabet.

\section{System Model and Problem Formulation}

\subsection{DMA Model}

DMA is a planar array consisting of multiple microstrips, each of
which contains a certain number of closely spaced radiating elements.
Moreover, every microstrip has a port at its end that is used to connect
it with a single \ac{RF} chain and transfer the transmitted/received
signals in between.\footnote{It should be noted that the number of RF chains for DMA-based transceivers
is significantly lower than the number of RF chains for fully digital
MIMO transceivers, which require an RF chain per radiating elements.
Even in comparison to hybrid MIMO transceivers that also use a reduced
number RF chains, DMA-based transceivers have lower hardware complexity,
as they do not need additional dedicated analog circuits (i.e., phase
shifters).} Signal propagation from the input port to the \mbox{$l$-th} element
in the $i$-th microstrip is modeled by the coefficient $h_{l,i}=e^{-\rho_{l,i}(\alpha_{i}+j\beta_{i})}$,
where $\alpha_{i}$ and $\beta_{i}$ denote the attenuation coefficient
and the wavenumber of the microstrip, respectively, and $\rho_{l,i}$
is the distance of this elements from the input port. After reaching
the radiating element, the propagation signal can be tuned by the
weight, $q_{l,i}$. Since we assume operating in a narrow-band system,
the feasible set of the weights of the DMA elements is approximated
by a Lorentzian constraint as
\begin{equation}
\mathcal{Q}=\{(j+e^{j\phi})/2\,\big|\,\phi\in[0,2\pi]\}.\label{eq:Q_def}
\end{equation}

For a DMA with $N_{d}$ microstrips and $N_{e}$ radiating elements
at each of them, the total number of radiating elements is $N=N_{d}N_{e}$.
The relation between the input signal $\mathbf{s}\in\mathbb{C}^{N_{d}\times1}$
and the transmitted (output) signal $\mathbf{t}\in\mathbb{C}^{N_{t}\times1}$
for this DMA is given by
\begin{equation}
\mathbf{t}=\mathbf{H}\mathbf{Q}\mathbf{s}
\end{equation}
where 
\begin{equation}
[\mathbf{Q}]_{(i-1)N_{e}+l,k}=\begin{cases}
q_{l,i}\in\mathcal{Q} & i=k\\
0 & i\neq k
\end{cases}\label{eq:Q_elem}
\end{equation}
is a matrix of size $N\times N_{d}$ that contains the tunable weights
of all radiating elements, and 
\begin{equation}
[\mathbf{H}]_{(i-1)N_{e}+l,(i-1)N_{e}+l}=h_{i,l}
\end{equation}
a diagonal matrix of size $N\times N$ that contains the propagation
coefficients of all radiating elements, and where $l=1,2,\dots,N_{e}$
and $i,k=1,2,\dots,N_{d}$.

\subsection{System Model}

We consider a point-to-point communication system, where a DMA-based
transmitter employs $N$ radiating elements for signal transmission
to a receiver with $N_{r}$ receive antennas. The transmit symbol
vector $\mathbf{x}_{m}\in\mathbb{C}^{N_{d}\times1}$ consists of $N_{d}$
elements selected from a discrete symbol alphabet of size $M$. This
 vector is precoded using a diagonal matrix $\mathbf{P}=\text{diag}(\mathbf{p})\in\mathbb{C}^{N_{d}\times N_{d}},$
so that the input signal of the DMA can be represented as $\mathbf{s}=\mathbf{P}\mathbf{x}_{m}$.
Therefore, the   signal vector at the receive antennas is given
by 
\begin{equation}
\mathbf{y}=\mathbf{G}\mathbf{H}\mathbf{Q}\mathbf{P}\mathbf{x}_{m}+\mathbf{n},\label{eq:sys_model}
\end{equation}
where $\mathbf{G}\in\mathbb{C}^{N_{r}\times N_{d}}$ is the channel
matrix. The noise vector $\mathbf{n}\in\mathbb{C}^{N_{r}\times1}$
consists of independent and identically distributed (i.i.d.) elements
that are distributed according to $\mathcal{CN}(0,\sigma^{2})$, where
$\sigma^{2}$ denotes the noise variance. We assume that the precoding
matrix $\mathbf{P}$ and the DMA jointly preserve the average power
of the transmitted signal, i.e., 
\begin{equation}
||\mathbf{H}\mathbf{Q}\mathbf{P}||^{2}=||\mathbf{H}\mathbf{Q}\text{diag}(\mathbf{p})||^{2}=N_{d}.\label{eq:P_constr}
\end{equation}

\subsection{Problem Formulation}

Based on the union bound method \cite{john2008digital}, an upper
bound to the BEP of the considered system  is given as 
\begin{align}
\text{BEP} & \le\frac{1}{N_{vec}\log_{2}N_{vec}}\sum_{\underset{m\neq n}{m,n=1}}^{N_{vec}}D(\mathbf{x}_{m},\mathbf{x}_{n})Q\left(\sqrt{\frac{F_{m,n}}{2\sigma^{2}}}\right)
\end{align}
where $N_{vec}=M^{N_{d}}$ is the number of different transmit symbol
vectors, $D(\mathbf{x}_{m},\mathbf{x}_{n})$ is the Hamming distance
(i.e., number of different bits) between bit representations of the
input vectors $\mathbf{x}_{m}$ and $\mathbf{x}_{n}$, and $F_{m,n}=\left\Vert \mathbf{G}\mathbf{H}\mathbf{Q}\mathbf{P}(\mathbf{x}_{m}-\mathbf{x}_{n})\right\Vert ^{2}=\left\Vert \mathbf{G}\mathbf{H}\mathbf{Q}\mathbf{P}\triangle\mathbf{x}_{m,n}\right\Vert ^{2}$
is the Euclidean distance between the received signal vectors $\mathbf{G}\mathbf{H}\mathbf{Q}\mathbf{P}\mathbf{x}_{m}$
and $\mathbf{G}\mathbf{H}\mathbf{Q}\mathbf{P}\mathbf{x}_{n}$.

From the previous expression and the constraints \eqref{eq:Q_def},
\eqref{eq:Q_elem} and \eqref{eq:P_constr}, the BEP optimization
problem can be formulated as 
\begin{subequations}
\begin{align}
\text{\ensuremath{\underset{\mathbf{Q},\mathbf{p}}{\minimize}}	} & f(\mathbf{Q},\mathbf{p})=\sum_{\underset{m\neq n}{m,n=1}}^{N_{vec}}D(\mathbf{x}_{m},\mathbf{x}_{n})Q\left(\sqrt{\frac{F_{m,n}}{2\sigma^{2}}}\right)\label{eq:obj_fun}\\
\st\text{	} & [\mathbf{Q}]_{(i-1)N_{e}+l,k}=\begin{cases}
q_{l,i}\in\mathcal{Q} & i=k\\
0 & i\neq k
\end{cases}\label{eq:phi-modulus}\\
 & ||\mathbf{H}\mathbf{Q}\text{diag}(\mathbf{p})||^{2}=N_{d}.
\end{align}
\label{eq:opt_prob}
\end{subequations}
 for $l=1,2,\dots,N_{e}$ and $i,k=1,2,\dots,N_{d}$.

\section{Proposed Optimization Method}

To solve \eqref{eq:opt_prob}, we propose an efficient PGM, which
updates $\mathbf{Q}$ and $\mathbf{p}$ according to
\begin{subequations}
\begin{align}
\mathbf{Q}^{(n+1)} & =P_{Q}(\mathbf{Q}^{(n)}-\mu_{1}\mathbf{M}\varodot\nabla_{\mathbf{\mathbf{Q}}}f(\mathbf{Q}^{(n)},\mathbf{p}^{(n)}))\label{eq:Update_Q}\\
\mathbf{p}^{(n+1)} & =P_{P}(\mathbf{p}^{(n)}-\mu_{2}\nabla_{\mathbf{p}}f(\mathbf{Q}^{(n+1)},\mathbf{p}^{(n)}))\label{eq:Update_p}
\end{align}
\end{subequations}
where $\nabla_{\mathbf{p}}f(\mathbf{p},\mathbf{Q})$ and $\nabla_{\mathbf{\mathbf{Q}}}f(\mathbf{p},\mathbf{Q})$
are the gradients of $f(\mathbf{p},\mathbf{Q})$ with respect to $\mathbf{p}^{*}$
and $\mathbf{Q}^{*}$, receptively, and $\mu_{1}$ and $\mu_{2}$
are the appropriate step sizes. In addition, $\varodot$ denotes the
element-wise product. The elements of the matrix $\mathbf{M}$ are
given as\footnote{The positions of zero elements in $\mathbf{M}$ correspond to the
positions of zero elements in $\mathbf{Q}$ and all other elements
in $\mathbf{M}$ are equal one. Hence, the positions of zero and non-zero
elements in $\mathbf{M}\varodot\nabla_{\mathbf{\mathbf{Q}}}f(\mathbf{p}^{(n)},\mathbf{Q}^{(n)})$
are the same as in $\mathbf{Q}$, which is necessary for the correct
work of the proposed algorithm.}
\begin{equation}
[\mathbf{M}]_{(i-1)N_{e}+l,k}=\begin{cases}
1 & i=k\\
0 & i\neq k
\end{cases}
\end{equation}
for $l=1,2,\dots,N_{e}$ and $i,k=1,2,\dots,N_{d}$. Furthermore,
$P_{P}(\cdot)$ and $P_{Q}(\cdot)$ denote the projection operations.
The involved  gradients are provided in the next lemma.
\begin{lem}
The gradient of $f(\mathbf{Q},\mathbf{p})$ w.r.t. $\mathbf{Q}^{*}$
and $\mathbf{p}^{*}$ are given by
\begin{gather}
\nabla_{\mathbf{Q}}f(\mathbf{Q},\mathbf{p})=-\frac{1}{4\sqrt{2\pi}\sigma^{2}}\mathbf{H}^{H}\mathbf{G}^{H}\mathbf{G}\mathbf{H}\mathbf{Q}\mathbf{P}\sum_{\underset{m\neq n}{m,n=1}}^{N_{vec}}D(\mathbf{x}_{m},\mathbf{x}_{n})\nonumber \\
\times e^{-\frac{F_{m,n}}{4\sigma^{2}N_{d}}}\left(\frac{F_{m,n}}{2\sigma^{2}}\right)^{-1/2}\triangle\mathbf{x}_{m,n}\triangle\mathbf{x}_{m,n}^{H}\mathbf{P}^{H}\label{eq:Grad_Q}
\end{gather}
\vspace{-10mm}

\begin{gather}
\nabla_{\mathbf{p}}f(\mathbf{Q},\mathbf{p})=-\frac{1}{4\sqrt{2\pi}\sigma^{2}}\vect_{d}\biggl(\mathbf{Q}^{H}\mathbf{H}^{H}\mathbf{G}^{H}\mathbf{G}\mathbf{H}\mathbf{Q}\mathbf{P}\nonumber \\
\times\sum_{\underset{m\neq n}{m,n=1}}^{N_{vec}}D(\mathbf{x}_{m},\mathbf{x}_{n})e^{-\frac{F_{m,n}}{4\sigma^{2}N_{d}}}\left(\frac{F_{m,n}}{2\sigma^{2}}\right)^{-1/2}\triangle\mathbf{x}_{m,n}\triangle\mathbf{x}_{m,n}^{H}\biggl)\label{eq:grad_P}
\end{gather}
where $\vect_{d}(\mathbf{X})$ denotes a column-vector whose elements
are the diagonal elements of $\mathbf{X}$.
\end{lem}
\begin{IEEEproof}
Differentiating the objective function, we obtain
\begin{gather}
\text{d}f(\mathbf{Q},\mathbf{p})=\sum_{\underset{m\neq n}{m,n=1}}^{N_{vec}}D(\mathbf{x}_{m},\mathbf{x}_{n})\text{d}Q\left(\sqrt{\frac{F_{m,n}}{2\sigma^{2}N_{d}}}\right)\nonumber \\
\stackrel{(a)}{=}-\frac{1}{4\sqrt{2\pi}\sigma^{2}}\sum_{\underset{m\neq n}{m,n=1}}^{N_{vec}}D(\mathbf{x}_{m},\mathbf{x}_{n})e^{-\frac{F_{m,n}}{4\sigma^{2}}}\left(\frac{F_{m,n}}{2\sigma^{2}}\right)^{-1/2}\text{d}F_{m,n}\label{eq:grad_gen}
\end{gather}
where in $(a)$, we used the identity $\text{d}Q(x)=-\frac{1}{\sqrt{2\pi}}e^{-\frac{x^{2}}{2}}\text{d}x.$

Furthermore, we have
\begin{gather}
\text{d}F_{m,n}=\text{d}||\mathbf{G}\mathbf{H}\mathbf{Q}\mathbf{P}\triangle\mathbf{x}_{m,n}||^{2}\nonumber \\
=\text{d}\tr\left(\mathbf{G}\mathbf{H}\mathbf{Q}\mathbf{P}\triangle\mathbf{x}_{m,n}\triangle\mathbf{x}_{m,n}^{H}\mathbf{P}^{H}\mathbf{Q}^{H}\mathbf{H}^{H}\mathbf{G}^{H}\right)\nonumber \\
=\tr(\mathbf{P}\triangle\mathbf{x}_{m,n}\triangle\mathbf{x}_{m,n}^{H}\mathbf{P}^{H}\mathbf{Q}^{H}\mathbf{H}^{H}\mathbf{G}^{H}\mathbf{G}\mathbf{H}\text{d}\mathbf{Q}\nonumber \\
+\mathbf{H}^{H}\mathbf{G}^{H}\mathbf{G}\mathbf{H}\mathbf{Q}\mathbf{P}\triangle\mathbf{x}_{m,n}\triangle\mathbf{x}_{m,n}^{H}\mathbf{P}^{H}\text{d}\mathbf{Q}^{H}).\label{eq:dif_P}
\end{gather}

Substituting \eqref{eq:dif_P} into \eqref{eq:grad_gen}, we obtain
\eqref{eq:Grad_Q}. Similarly, we can prove \eqref{eq:grad_P}.
\end{IEEEproof}
Since the wights of the DMA radiating elements are constrained as
stated in \eqref{eq:Q_def}, the projection $P_{Q}(\cdot)$ is performed
for each non-zero element of $\mathbf{Q}$ according to 
\begin{equation}
[\bar{\mathbf{Q}}]_{i,j}=\underset{q\in\mathcal{Q}}{\arg\min}\left|[\mathbf{Q}]_{i,j}-q\right|.
\end{equation}

Moreover, the constraint \eqref{eq:P_constr} implies that the projection
$P_{P}(\cdot)$ can be written as 
\begin{equation}
\bar{\mathbf{p}}=\sqrt{N_{d}}\mathbf{p}/||\mathbf{H}\mathbf{Q}\text{diag}(\mathbf{p})||.
\end{equation}

Finally, we need to find the step sizes in \eqref{eq:Update_Q} and
\eqref{eq:Update_p} to ensure the convergence of the proposed algorithm.
Utilizing the Armijo-Goldstein backtracking line search, we determine
the appropriate step sizes $\mu_{1}$ and $\mu_{2}$ at each iteration.
For $\mu_{\text{init}}>0$, a small constant $\delta>0$ and $\rho\in(0,1)$,
the step size $\mu_{1}$ is $\mu_{\text{init}}\rho^{l_{n}}$, where
$l_{n}$ is the smallest nonnegative integer such that
\begin{equation}
\!\!\!f(\mathbf{Q}^{(n+1)},\mathbf{p}^{(n)})\le f(\mathbf{Q}^{(n)},\mathbf{p}^{(n)})-\delta||\mathbf{Q}^{(n+1)}-\mathbf{Q}^{(n)}||^{2}.
\end{equation}
Similarly, the step size $\mu_{2}$ can be determined.

\section{Convergence and Complexity Analysis}

\subsection{Convergence Analysis}

In this section, we provide the convergence analysis of the PGM, which
is used in this paper. Let us recall the following inequality for
any function $g(x)$ which is L-smooth:
\begin{equation}
g(y)\le g(x)+\left\langle \nabla_{x}g(x),y-x\right\rangle +\frac{L}{2}||y-x||^{2}
\end{equation}
where $L$ is the Lipschitz constant. Implementing this identity for
the optimization of $\mathbf{p}$, we obtain
\begin{gather}
f(\mathbf{Q}^{(n+1)},\mathbf{p}^{(n+1)})\le f(\mathbf{Q}^{(n+1)},\mathbf{p}^{(n)})+\nonumber \\
\left\langle \nabla_{\mathbf{p}}f(\mathbf{Q}^{(n+1)},\mathbf{p}^{(n)}),\mathbf{p}^{(n+1)}-\mathbf{p}^{(n)}\right\rangle +\frac{L_{2}}{2}||\mathbf{p}^{(n+1)}-\mathbf{p}^{(n)}||^{2}\label{eq:fP_L}
\end{gather}
where $L_{2}$ is the Lipschitz constant of $\nabla_{\mathbf{p}}f(\mathbf{Q}^{(n+1)},\mathbf{p}^{(n)})$
and $\left\langle \mathbf{x},\mathbf{y}\right\rangle =\mathcal{\mathfrak{R}}\{\mathbf{x}^{H}\mathbf{y}\}$.
The projection of $\mathbf{p}^{(n+1)}$ onto the feasible set can
be expressed as
\begin{gather*}
\mathbf{p}^{(n+1)}=\underset{\mathbf{p}}{\arg\min}\:||\mathbf{p}-\mathbf{p}^{(n)}+\mu_{2}\nabla_{\mathbf{p}}f(\mathbf{Q}^{(n+1)},\mathbf{p}^{(n)})||^{2}=\\
\underset{\mathbf{p}}{\arg\min}\left\langle \nabla_{\mathbf{p}}f(\mathbf{Q}^{(n+1)},\mathbf{p}^{(n)}),\mathbf{p}-\mathbf{p}^{(n)}\right\rangle +\frac{1}{2\ensuremath{\mu_{2}}}||\mathbf{p}-\mathbf{p}^{(n)}||^{2}.
\end{gather*}
 Thus, it is obvious that
\begin{multline}
\!\!\!\!\!\!\!\!\!\left\langle \nabla_{\mathbf{p}}f\mathbf{Q}^{(n+1)},\mathbf{p}^{(n)}),\mathbf{p}^{(n+1)}-\mathbf{p}^{(n)}\right\rangle +\frac{1}{2\ensuremath{\mu_{2}}}||\mathbf{p}^{(n+1)}-\mathbf{p}^{(n)}||^{2}\leq\\
\!\!\!\left.\left\langle \nabla_{\mathbf{p}}f(\mathbf{Q}^{(n+1)},\mathbf{p}^{(n)}),\mathbf{p}-\mathbf{p}^{(n)}\right\rangle +\frac{1}{2\ensuremath{\mu_{2}}}||\mathbf{p}-\mathbf{p}^{(n)}||^{2}\right|_{\mathbf{p}=\mathbf{p}^{(n)}}=0.\label{eq:fP_ni}
\end{multline}
Combining \eqref{eq:fP_L} and \eqref{eq:fP_ni}, we have
\begin{gather}
f(\mathbf{Q}^{(n+1)},\mathbf{p}^{(n+1)})\le f(\mathbf{Q}^{(n+1)},\mathbf{p}^{(n)})-\nonumber \\
\left(1/(2\ensuremath{\mu_{2}})-L_{2}/2\right)||\mathbf{p}^{(n+1)}-\mathbf{p}^{(n)}||^{2}.\label{eq:fP}
\end{gather}
Similarly for the optimization of $\mathbf{Q}$, we can write
\begin{gather}
f(\mathbf{Q}^{(n+1)},\mathbf{p}^{(n)})\le f(\mathbf{Q}^{(n)},\mathbf{p}^{(n)})-\nonumber \\
\left(1/(2\ensuremath{\mu_{1}})-L_{1}/2\right)||\mathbf{Q}^{(n+1)}-\mathbf{Q}^{(n)}||^{2}\label{eq:fQ}
\end{gather}
where $L_{1}$ is the Lipschitz constant of $\nabla_{\mathbf{Q}}f(\mathbf{p}^{(n+1)},\mathbf{Q}^{(n)})$.

After combining \eqref{eq:fP} and \eqref{eq:fQ}, we obtain 
\begin{gather}
\!\!\!\!\!f(\mathbf{Q}^{(n+1)},\mathbf{p}^{(n+1)})\le f(\mathbf{Q}^{(n)},\mathbf{p}^{(n)})-\left(1/(2\ensuremath{\mu_{1})}-L_{1}/2\right)\nonumber \\
\times||\mathbf{Q}^{(n+1)}-\mathbf{Q}^{(n)}||^{2}-\left(1/(2\ensuremath{\mu_{2}})-L_{2}/2\right)||\mathbf{p}^{(n+1)}-\mathbf{p}^{(n)}||^{2}.
\end{gather}

Since the backtracking line search procedure ensures that $\frac{1}{\ensuremath{\mu_{1}}}>L_{1}$
and $\frac{1}{\ensuremath{\mu_{2}}}>L_{2}$, the sequence $\{f(\mathbf{Q}^{(n)},\mathbf{p}^{(n)})\}$
is strictly decreasing. Moreover, $f(\mathbf{Q}^{(n)},\mathbf{p}^{(n)})$
is bounded from below due to the continuity $f(\mathbf{Q},\mathbf{p})$
and the compactness of the feasible set. Therefore, the sequence $\{(\mathbf{Q}^{(n)},\mathbf{p}^{(n)})\}$
is indeed convergent. In addition, according to \cite{bolte2014proximal},
it can be easily shown that this sequence converges to a stationary
point.

\subsection{Complexity Analysis}

The computational complexity of the proposed algorithm is obtained
by counting the required number of complex multiplications.  The
computational complexity of calculating the gradient $\nabla_{\mathbf{Q}}f(\mathbf{Q},\mathbf{p})$
is equal to $\mathcal{O}(NN_{r}N_{d}+N_{vec}^{2}N_{d}^{2})$. Furthermore,
$\mathcal{O}(NN_{r}+N_{vec}^{2}N_{r}N_{d})$ multiplications are required
to obtain $f(\mathbf{Q}^{(n+1)},\mathbf{p}^{(n)})$. Hence, the complexity
of optimizing the matrix $\mathbf{Q}$ is $\mathcal{O}(NN_{r}N_{d}+N_{vec}^{2}N_{d}^{2}+I_{Q}(NN_{r}+N_{vec}^{2}N_{r}N_{d}))$,
where $I_{Q}$ is the number of line search steps. Similarly, the
complexity of computing the gradient $\nabla_{\mathbf{p}}f(\mathbf{Q},\mathbf{p})$
is $\mathcal{O}(N_{vec}^{2}N_{d}^{2})$ and $\mathcal{O}(N_{vec}^{2}N_{r}N_{d})$
multiplications is need to obtain $f(\mathbf{Q}^{(n+1)},\mathbf{p}^{(n+1)})$.
Therefore, the complexity of optimizing $\mathbf{p}$ is $\mathcal{O}(N_{vec}^{2}N_{d}^{2}+I_{P}N_{vec}^{2}N_{r}N_{d})$,
where $I_{P}$ is the number of line search steps. Hence, the~complexity
of one iteration of the proposed algorithm is equal to
\begin{align}
C & =\mathcal{O}(NN_{r}N_{d}+N_{vec}^{2}N_{d}^{2}+I_{Q}(NN_{r}+N_{vec}^{2}N_{r}N_{d})\nonumber \\
 & +I_{P}N_{vec}^{2}N_{r}N_{d}).
\end{align}

\section{Simulation Results}

In the following simulations, the channel matrices are~modeled according
to the sparsely-scattered channel model as 
\begin{equation}
\mathbf{G}=\sqrt{\frac{NN_{r}}{L}}\sum_{l=1}^{L}\alpha_{l}\mathbf{a}_{\text{R}}(\phi_{r})\mathbf{a}_{\text{T}}(\theta_{t},\phi_{t})^{H}
\end{equation}
where $L$ is the number of paths and $\alpha_{l}=\lambda/(4\pi d_{l})$
is the complex gain of the $l$-th path, whose length is denoted as
$d_{l}$ \cite{zhang2023channel,el2014spatially}. Moreover, $\phi_{t}(\theta_{t})$
are the azimuth (elevation) angles of departure, and $\phi_{r}$ is
the azimuth angle of arrival. The array response vectors of the transmit
DMA, $\mathbf{a}_{\text{T}}(\theta_{t},\phi_{t})$, and the receive
ULA, $\mathbf{a}_{\text{R}}(\phi_{r})$, are calculated according
to \cite{zhang2023channel}.

The simulation setup considered in this paper assumes a frequency
equal to $f=28\thinspace\mathrm{GHz}$, whose corresponding wavelength
is $\lambda=1.07\thinspace\mathrm{cm}$. The transmit signal power
is $\mathbb{E}\{||\mathbf{x}_{m}||^{2}\}=1\,\mathrm{W}$ and the noise
variance (power) is $\sigma^{2}=-105\thinspace\mathrm{dB}$. Each
microstrip consists $N_{e}=10$ elements and the separation between
the adjacent radiation element is $\lambda/2$. As in \cite{wang2020dynamic},
we use $\alpha_{i}=0.6\thinspace\text{m}^{-1}$ and $\beta_{i}=827.67\thinspace\text{m}^{-1}$
to represent the propagation inside all microstrips. The angles $\phi_{t}$,
$\theta_{t}$ and $\phi_{r}$ are randomly sampled values from the
interval $[\pi/6,5\pi/6]$. Similarly, the path length $d_{l}$ is
randomly selected from the interval $[10\,\text{m},40\,\text{m}]$.
For the proposed algorithm, all step sizes are initially set to 1000,
$\delta=10^{-3}$ and $\rho=1/2$. To speed up the convergence of
the proposed algorithm, all step sizes are reinitialized after every
50 iterations. The initial values of the optimization variables $\mathbf{p}$
and $\mathbf{Q}$ are randomly generated. All results are averaged
over 500 independent channel realizations.

\begin{figure}[t]
\centering{}\includegraphics[width=8.85cm]{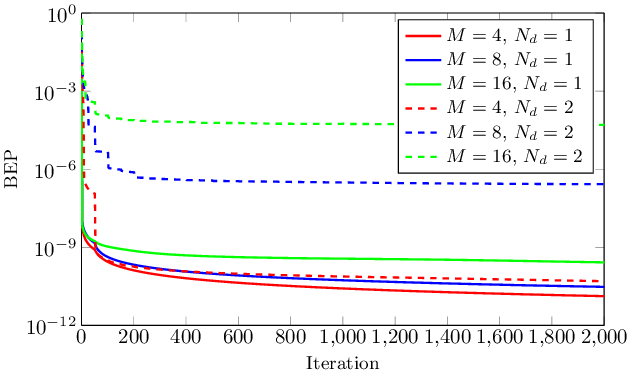}\caption{BEP for different size of symbol alphabet ($M$) and number of microstrips
($N_{d}$). \label{fig:BEP-vs-M}}
\end{figure}
In Fig. \ref{fig:BEP-vs-M}, we show the BEP for different size of
symbol alphabet $M$ and number of micostrips $N_{d}$. In the considered
scenario, the number of the receive antennas $N_{r}$ is equal to
the number of microstrips $N_{d}$. We can observe that the BEP is
increased with $N_{d}$, since the average transmit power is always
the same regardless of the number of microstrips. This BEP enlargement
changes with $M$, so that it is relatively small for $M=4$ and is
about five orders of magnitude for $M=16$. This phenomenon can be
explained by the following argumentation. The BEP primarily depends
on the Euclidean distances between the received signal vectors, $F_{m,n}$,
especially the smallest ones. Moreover, the number of these smallest
distances $F_{m,n}$, that have the most influence on the BEP, is
generally increased with $M$ and $N_{d}$. As a result, the BEP is
more substantially increased with the number of microstrips when the
size of symbol alphabet is large.

\section{Conclusion}

In this paper, we have studied the BEP optimization of DMA-based transmission
in a point-to-point MIMO communication system and proposed an optimization
algorithm. This algorithm is based on the PGM for which we derived
the closed form gradient expressions. Moreover, we proved the convergence
of this algorithm and presented its computational complexity. Simulation
results showed that the BEP is more sensitive to the number of microstrips
in DMA-based systems with a larger size of symbol alphabet.

\bibliographystyle{IEEEtran}
\bibliography{IEEEabrv,references,IEEEexample}

\end{document}

%% file: acro.tex
\acrodef{ADC}{analog-to-digital converter}
\acrodef{AO}{alternating optimization}
\acrodef{APGM}{alternating projected gradient method}
\acrodef{APM}{accelerated proximal gradient method}
\acrodef{ASP}{antenna separation product}
\acrodef{AWGN}{additive white Gaussian noise}
\acrodef{BC}{broadcast channel}
\acrodef{BCM}{block coordinate maximization}
\acrodef{BEP}{bit error probability}
\acrodef{BER}{bit error rate}
\acrodef{BF-MIMO}[BF\mbox{-}MIMO]{beamforming MIMO}
\acrodef{BF}{beamforming}
\acrodef{BS}{base station}
\acrodef{bpcu}{bits per channel use}
\acrodef{CP}{cyclic prefix}
\acrodef{CR}{cutoff rate}
\acrodef{CSI}{channel state information}
\acrodef{CSIR}{channel state information at RX}
\acrodef{SSK}{space shift keying}
\acrodef{CSIT}{channel state information at TX}
\acrodef{DCMC}{discrete\mbox{-}input continuous\mbox{-}output memoryless channel}
\acrodef{DFT}{discrete Fourier transform}
\acrodef{DL-TR-GSM}{dual-layered transmit-receive \acl{GSM}}
\acrodef{DLT}{dual-layered transmission}
\acrodef{DMA}{dynamic metasurface antenna}
\acrodef{DOA}{direction of arrival}
\acrodef{DPC}{dirty paper coding}
\acrodef{EGC}{equal gain combining}
\acrodef{EM}{electromagnetic}
\acrodef{EVD}{eigenvalue decomposition}
\acrodef{FPGA}{field programmable gate array}
\acrodef{FSPL}{free space path loss}
\acrodef{FFT}{fast Fourier transform}
\acrodef{FDE}{frequency domain equalization}
\acrodef{GRSM}{generalized \acl{RSM}}
\acrodef{GSM}{generalized \acl{SM}}
\acrodef{HMIMO}{holographic MIMO}
\acrodef{IFFT}{invserse fast Fourier transform}
\acrodef{ICI}{inter-channel interference}
\acrodef{iid}[i.i.d.]{independent and identically distributed}
\acrodef{IQ}{in\mbox{-}phase and quadrature}
\acrodef{ISI}{intersymbol interference}
\acrodef{ISI-free}[ISI\mbox{-}free]{intersymbol interference free}
\acrodef{LIS}{large intelligent surface}
\acrodef{LOS}{line\mbox{-}of\mbox{-}sight}
\acrodef{KKT}{Karush\mbox{-}Kuhn\mbox{-}Tucker} 
\acrodef{MAC}{multiple-access channel}
\acrodef{mmWave}{millimeter-wave}
\acrodef{MI}{mutual information}
\acrodef{MIMO}{multiple\mbox{-}input multiple\mbox{-}output}
\acrodef{mMIMO}{massive MIMO}
\acrodef{MISO}{multiple\mbox{-}input single\mbox{-}output}
\acrodef{ML}{maximum likelihood}
\acrodef{MRC}{maximal ratio combining}
\acrodef{MMSE}{minimum mean square error}
\acrodef{MU-TR-GSM}{multiuser transmit-receive  \acl{GSM} }
\acrodef{NCSIT}{no channel state information at TX}
\acrodef{NLOS}{non\mbox{-}\acs{LOS}} 
\acrodef{NOMA}{non-orthogonal multiple access}
\acrodef{OFDM}{orthogonal frequency division multiplexing}
\acrodef{OFDMA}{orthogonal frequency division multiple access}
\acrodef{PA}{power amplifier}
\acrodef{PAE}{power added efficiency}
\acrodef{PAPR}{peak\mbox{-}to\mbox{-}average power ratio}
\acrodef{PDF}{probability density function}
\acrodef{PEP}{pairwise error probability}
\acrodefplural{PEP}{pairwise error probabilities}
\acrodef{PGM}{projected gradient method}
\acrodef{PMP}{probability mass function}
\acrodef{PSM}{precoding-aided spatial modulation}
\acrodef{QSM}{quadrature spatial modulation}
\acrodef{RC}{reorganization computation}
\acrodef{RF}{radio frequency}
\acrodef{RHS}{right-hand side}
\acrodef{RIS}{reconfigurable intelligent surface}
\acrodef{RSM}{receive spatial modulation}
\acrodef{RX}{receiver}
\acrodef{SEP}{symbol error probability}
\acrodef{SER}{symbol error rate}
\acrodef{SIC}{successive interference cancellation}
\acrodef{SIM}{stacked intelligent metasurface}
\acrodef{SINR}{signal-to-interference-plus-noise ratio}
\acrodef{SISO}{single-input single-output}
\acrodef{SM}{spatial modulation}
\acrodef{SMX-MIMO}[SMX\mbox{-}MIMO]{spatial multiplexing MIMO}
\acrodef{SMX}{spatial multiplexing}
\acrodef{SNR}{signal-to-noise ratio}
\acrodef{SC}{single carrier}
\acrodef{SVD}{singular value decomposition}
\acrodef{SPST}{single pole single-throw}
\acrodef{SU}{secondary user}
\acrodef{TDE}{time domain equalization}
\acrodef{TX}{transmitter}
\acrodef{ULA}{uniform linear array}
\acrodef{URA}{uniform rectangular array}
\acrodef{VGA}{variable gain amplifier}
\acrodef{wrt}[w.r.t.]{with respect to}
\acrodef{ZF}{zero-forcing}
\acrodef{ZMCG}{zero-mean complex Gaussian}